\documentclass[%
 reprint,
superscriptaddress,
 amsmath,amssymb,
 aps,
 prl,
]{revtex4-1}

\usepackage{graphicx}
\usepackage{dcolumn, braket}
\usepackage{bm}
\usepackage{xcolor}

\begin{document}

\author{Michele~Pizzochero}
\email{mpizzochero@g.harvard.edu}
\affiliation{School of Engineering and Applied Sciences, Harvard University, Cambridge, MA 02138, United States}

\author{Nikita V. Tepliakov}
\email{n.tepliakov20@imperial.ac.uk}
\affiliation{Departments of Materials and Physics, Imperial College London, London SW7 2AZ, United Kingdom}
\affiliation{The Thomas Young Centre for Theory and Simulation of Materials, \\ Imperial College London, London SW7 2AZ, United Kingdom}

\author{Johannes~Lischner}
\affiliation{Departments of Materials and Physics, Imperial College London, London SW7 2AZ, United Kingdom}
\affiliation{The Thomas Young Centre for Theory and Simulation of Materials, \\ Imperial College London, London SW7 2AZ, United Kingdom}

\author{Arash~A.~Mostofi}
\affiliation{Departments of Materials and Physics, Imperial College London, London SW7 2AZ, United Kingdom}
\affiliation{The Thomas Young Centre for Theory and Simulation of Materials, \\ Imperial College London, London SW7 2AZ, United Kingdom}

\author{Efthimios~Kaxiras}
\affiliation{Department of Physics, Harvard University, Cambridge, MA 02138, United States}
\affiliation{School of Engineering and Applied Sciences, Harvard University, Cambridge, MA 02138, United States}

\title{One-dimensional magnetic conduction channels across zigzag graphene nanoribbon/hexagonal boron nitride heterojunctions}

\newpage
\begin{abstract} 
We examine the electronic structure of recently fabricated in-plane heterojunctions of zigzag graphene nanoribbons embedded in hexagonal boron nitride. We focus on hitherto unexplored interface configurations in which both edges of the nanoribbon are bonded to the same chemical species, either boron or nitrogen atoms. \textcolor{black}{Using \emph{ab initio} and mean-field Hubbard model calculations, we} reveal the emergence of one-dimensional magnetic conducting channels \textcolor{black}{at these interfaces}. These channels originate from the energy shift of the magnetic interface states that is induced by charge transfer between the nanoribbon and hexagonal boron nitride.  We further address the response of these heterojunctions to external electric and magnetic fields, demonstrating the tunability of energy and spin splittings in the \textcolor{black}{electronic} structure. Our findings establish \textcolor{black}{that} zigzag graphene nanoribbon/hexagonal boron nitride heterojunctions \textcolor{black}{are} a suitable platform for exploring and engineering spin transport in the atomically thin limit, with potential applications in integrated spintronic devices.
 \end{abstract}

\maketitle

Graphene nanoribbons (GNRs) are a class of quasi-one dimensional semiconductors consisting of few-nanometer-wide strips of $sp^2$-bonded carbon atoms \cite{Yazyev2013, Zongping2020}. By virtue of their width-dependent band gaps \cite{Son2006a, Han2007, Chen2013, Pizzochero2020a}, high charge-carrier mobility \cite{Wang2021a, Baringhaus2014}, multifaceted structure-property relationships \cite{Cao2017, Groning2018, Rizzo2018, Arnold2022, Tepliakov2023}, and facile integration in short-channel field-effect transistors \cite{Bennett2013, Llinas2017, BorinBarin2019}, they have emerged as suitable building blocks for the realization of post-silicon device concepts \cite{Wang2021a}. 

Zigzag-edged graphene nanoribbons (ZGNRs) are peculiar owing to their inherent $\pi$-electron magnetism \cite{Yazyev2010, Ruffieux2016}. A Stoner instability originating from a van Hove singularity at the Fermi level \cite{Fujita1996} drives the formation of spin-polarized edge states \cite{Blackwell2021, Brede2023}, with the local magnetic moments on opposite edges favoring an antiparallel orientation. This leads to a spin-0 antiferromagnetic ground state, in agreement with Lieb's theorem for bipartite lattices at half filling \cite{Lieb1989}. The intrinsic magnetism of ZGNRs has been predicted to persist at room temperature \cite{Magda2014} and is tunable via electric  \cite{Son2006b, Kan2007} or strain fields \cite{Zhang2017, Hu2012}, as well as defect engineering \cite{Pizzochero2021, Pizzochero2022}. The advantageous combination of versatile magnetism with weak spin-orbit and hyperfine interactions \cite{Han2014, Slota2018}, the primary mechanisms causing spin relaxation and decoherence, renders ZGNRs an \textcolor{black}{interesting} platform for \textcolor{black}{future} spin-logic operations \cite{Yazyev2008}.

Recently, a two-step growth approach to laterally incorporate ZGNRs into hexagonal boron nitride ($h$BN) has been devised \cite{Chen2017, Wang2021b}. In the first step, nanotrenches in the surface layer of $h$BN are etched along the zigzag crystallographic orientation using nickel nanoparticle-catalyzed cutting. In the second step, chemical vapor deposition is used to fill these trenches with carbon atoms. This approach yields atomically precise, in-plane heterojunctions of ZGNRs of defined width seamlessly embedded in a continuous $h$BN matrix. \textcolor{black}{Previous computational work has}  established the occurrence of a Dirac half-semimetallic phase in these heterojunctions, where one spin orientation is semimetallic while the other is semiconducting \cite{Pruneda2010, Zeng2016, ding2009electronic, Tepliakov2023b}. This intriguing phase \textcolor{black}{has proven} to be robust against structural disorder  and \textcolor{black}{can be} modulated by charge doping \cite{Tepliakov2023b}, thus holding promise for antiferromagnetic spintronics and related applications \cite{Xiao2012, zhang2023ferrimagnets}.  However, \textcolor{black}{these} investigations have been limited to asymmetric configurations of the interfaces between the constituent materials, in which one edge of the nanoribbon is terminated by boron atoms and the other by nitrogen atoms of the surrounding hexagonal boron nitride. \textcolor{black}{Conversely, symmetric configurations where both edges of the embedded nanoribbon are terminated by the same chemical species}, either boron or nitrogen atoms, have received little attention. This issue is \textcolor{black}{important for} a complete understanding of ZGNR/$h$BN heterojunctions and \textcolor{black}{their possible use} in ultrathin nanocircuitry.

\begin{figure}[t]
    \centering
\includegraphics[width=1\columnwidth]{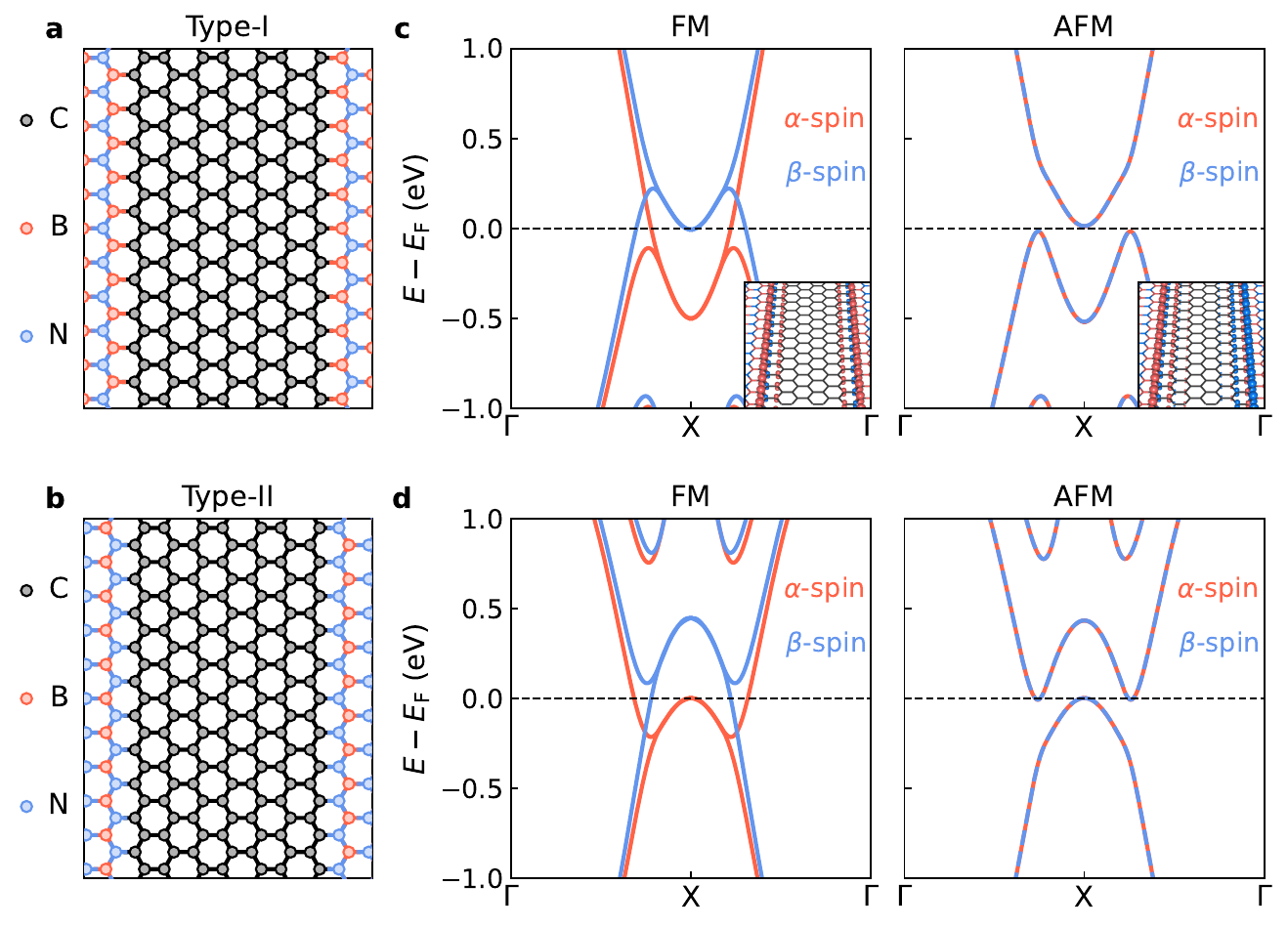}
    \caption{\textbf{Electronic structure of ZGNR/$h$BN heterojunctions from \emph{ab initio} calculations.} 
    \textbf{a}, Atomic structure of a type-I heterojunction, where the ZGNR embedded in $h$BN is terminated at both edges by boron atoms.
    \textbf{b}, Atomic structure of a type-II heterojunction, where the ZGNR embedded in $h$BN is terminated at both edges by nitrogen atoms.
    In panels \textbf{a} and \textbf{b}, black, red, and blue circles denote carbon, boron, and nitrogen atoms, respectively.
    \textbf{c}, Spin-resolved electronic band structures of a type-I heterojunction for a representative ZGNR of width 1.4 nm shown in panel \textbf{a} in the ferromagnetic (FM) and antiferromagnetic (AFM) phase. The insets show the spin density in the FM and AFM phases, where $\alpha$-spin and  $\beta$-spin contributions are shown as red and blue isosurfaces, respectively.
    \textbf{d}, Spin-resolved electronic band structures of a type-II heterojunction for a representative ZGNR of width 1.4 shown in panel \textbf{b} in the ferromagnetic (FM) and antiferromagnetic (AFM) phase.
    In panels \textbf{c} and \textbf{d}, $\alpha$-spin and  $\beta$-spin electronic bands are indicated as red and blue lines, respectively, and energies are referenced to the Fermi level.}
    \label{Fig1}
\end{figure}

\textcolor{black}{Here}, we bridge this gap by \textcolor{black}{considering} the effect of symmetric termination of the zigzag edges of the nanoribbon on the electronic structure of ZGNR/$h$BN  heterojunctions. By combining density functional theory and model \textcolor{black}{h}amiltonian calculations, we demonstrate the emergence of one-dimensional magnetic conducting channels at the interfaces. We further study the response of these heterojunctions to external electric and magnetic fields, revealing a significant tunability of the \textcolor{black}{electronic} structure in terms of energy,  spin, and momentum splittings. \textcolor{black}{We point out} similarities and differences with the well-studied asymmetric termination of the nanoribbon. Overall, our findings provide a comprehensive view of the electronic structure of ZGNR/$h$BN lateral heterojunctions from a theoretical perspective.

Our \emph{ab initio} calculations are based on spin-polarized density functional theory \cite{DFT}, adopting the generalized gradient approximation to the exchange-correlation functional proposed by Perdew, Burke, and Ernzerhof \cite{PBE}, as implemented in the \textsc{siesta} package \cite{SIESTA}. \textcolor{black}{Our computational models consist of one-dimensional graphene nanoribbons embedded in mirror-symmetric patches of hexagonal boron nitride.  This ensures that the edges of the ZGNR are terminated by the same chemical species of the surrounding $h$BN}. Further technical details of the \emph{ab initio} methodology are provided in Supporting Notes S1 \textcolor{black}{and S2}. Depending on the chemical environment of the carbon atoms at the edges of the ZGNR embedded in $h$BN, we distinguish three types of heterojunctions: (i) type-I, in which both nanoribbon edges are terminated by boron atoms; (ii) type-II, in which both edges are terminated by nitrogen atoms; (iii) type-III, in which one edge is terminated by boron atoms while the opposite edge is terminated by nitrogen atoms. The atomic structure of type-I and type-II heterojunctions is shown in \textcolor{black}{Fig.}\ 1(a) and \textcolor{black}{Fig.}\ 1(b), respectively, whereas a representative type-III heterojunction is \textcolor{black}{shown} in Supporting Note S3.

We begin our \emph{ab initio} investigation by examining a set of representative heterojunctions comprising zigzag graphene nanoribbons of width 1.4 nm. Of the three heterojunctions considered, type-I is the most stable one, lying  \textcolor{black}{0.17 eV/{\AA}} and \textcolor{black}{0.31 eV/{\AA}} per interface lower in energy than type-II and type-III heterojunctions, respectively, \textcolor{black}{when the same stoichiometry is considered}. Regardless of the edge termination, the incorporation of  ZGNRs into $h$BN preserves the $\pi$-electron magnetism. This results in a pair of one-dimensional arrays of magnetic moments, each of them of magnitude 0.23 $\mu$\textsubscript{B} per edge carbon atom, which are selectively localized at the interfaces between ZGNRs and the surrounding $h$BN matrix, as displayed in the insets of \textcolor{black}{Fig.}\ 1(c). 

The chemical environment in the vicinity of the edges of the ZGNR in the heterojunction strongly affects the relative stability between the ferromagnetic (i.e., parallel alignment of magnetic moments across the nanoribbon) and antiferromagnetic (i.e., antiparallel alignment of magnetic moments) \textcolor{black}{configurations}. On the one hand, in type-I and type-II heterojunctions, the ferromagnetic phase is more stable than the antiferromagnetic phase by 3.2 meV and 4.0 meV \textcolor{black}{per unit cell}, respectively. On the other hand, in the type-III heterojunction, the antiferromagnetic phase is more stable than the ferromagnetic phase by 2.4 meV \textcolor{black}{per unit cell}, qualitatively similar to hydrogen-terminated zigzag graphene nanoribbons where this value is 3.6 meV \cite{Tepliakov2023b}. 

In \textcolor{black}{Fig.}\ 1(c), we present the spin-resolved electronic band structure of the type-I and type-II heterojunctions in the ferromagnetic and antiferromagnetic phases. The corresponding results for the \textcolor{black}{non-magnetic} phase are given in Supporting Note S4. \textcolor{black}{On the one hand, ferromagnetic heterojunctions exhibit a metallic character.} \textcolor{black}{On the other hand, antiferromagnetic heterojunctions exhibit a semimetallic character, in which the valence and conduction bands touch the Fermi level at different points in the Brillouin zone.}  \textcolor{black}{These electronic and magnetic properties are quite insensitive to charge doping that may originate from, e.g., the underlying substrate, as we discuss in Supporting Notes S5 and S6}.  Our findings thus indicate that, upon incorporation into these heterojunctions, zigzag graphene nanoribbons act as one-dimensional magnetic and conducting channels across the layer of hexagonal boron nitride in which they are embedded. \textcolor{black}{This is surprising, since both ZGNRs and $h$BN are insulating in their free-standing form}. \textcolor{black}{The acquired (semi-)metallicity is present at all widths of the embedded ZGNR, as shown in Supporting Notes S7 \textcolor{black}{and S8},} in agreement with recent charge transport measurements performed on these heterojunctions, in which signatures of non-zero magnetic conductance at the Fermi level have been detected \cite{Wang2021a} \textcolor{black}{in wide ZGNR embedded in $h$BN}. \textcolor{black}{The (semi)metallic character of these heterojunctions} is different from type-III heterojunctions, which possess a vanishing density of states at the Fermi level and a Dirac half-semimetallic character where the insulating behavior in one spin orientation is accompanied by a Dirac semimetallic behavior in the other, as discussed in both Supporting Note S3 and prior studies \cite{Pruneda2010, Zeng2016, ding2009electronic, Tepliakov2023b}.  \textcolor{black}{In Supporting Note S9, we present the local density of states, proportional to the STM signal, of these heterojunctions, which is found to be localized along the edges of the embedded nanoribbon and to reach its maxima at the ZGNR/$h$BN interfaces. We thus suggest that STM imaging can effectively differentiate between pristine regions of $h$BN and those hosting ZGNRs and be used to precisely quantify the width of the embedded ZGNR.}

We determine the charge transfer that takes place at the interfaces between ZGNR and $h$BN in the heterojunctions through a Bader analysis \cite{Tang2009}. In type-I heterojunctions, the carbon atoms at the zigzag edges gain a negative charge of  $-0.96 \vert e \vert$  from boron atoms, while in type-II heterojunctions they gain a positive charge of $0.95 \vert e \vert$ from donating electrons to nitrogen atoms, in line with the difference in electronegativity between these chemical elements. The charge transfer is strictly localized at the edges of the ZGNR, with vanishing net atomic charges being found already at \textcolor{black}{the carbon atoms that are second nearest neighbor of the interface} sites.

\begin{figure}[t]
    \centering
\includegraphics[width=1\columnwidth]{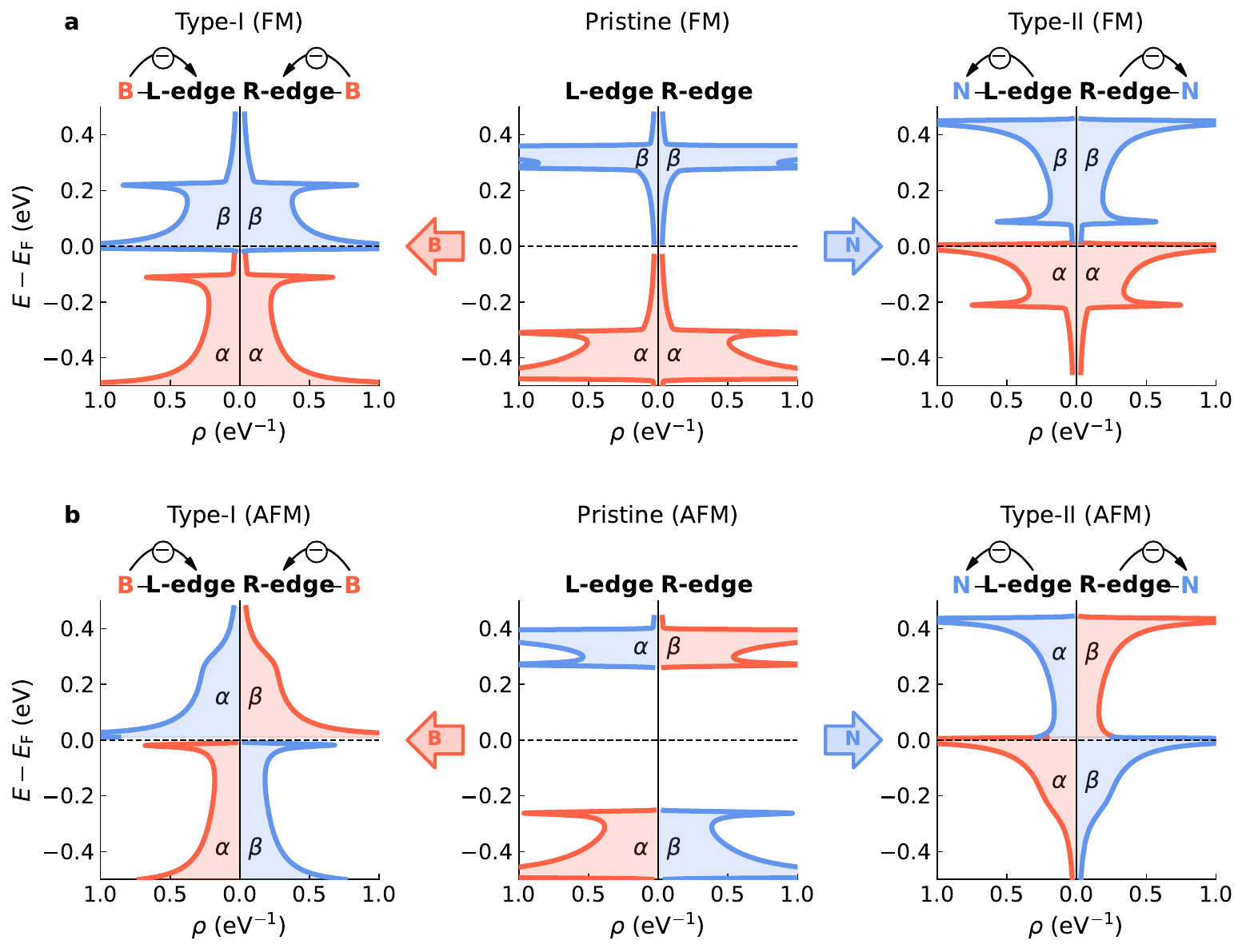}
    \caption{\textbf{Mechanism underlying the metallicity of ZGNR/$h$BN heterojunctions.}  \textbf{a}, Electronic density of states $\rho$ projected on the zigzag carbon atoms of the left (L-) and right (R-) edge of ZGNR in the ferromagnetic (FM) phase of a type-I ZGNR/$h$BN heterojunction (left panel), free-standing hydrogen-terminated ZGNR (central panel), and type-II ZGNR/$h$BN heterojunction (right panel).  \textbf{b}, Electronic density of states projected on the zigzag carbon atoms of the left (L-) and right (R-) edge of ZGNR in the antiferromagnetic (AFM) phase of a type-I ZGNR/$h$BN heterojunction (left panel), free-standing ZGNR (central panel), and type-II ZGNR/$h$BN heterojunction (right panel). The charge transfer induced from the boron (carbon) atom to the carbon (nitrogen) atom in type-I (type-II) heterojunctions is indicated by black arrows. $\alpha$-spin and  $\beta$-spin electronic bands are indicated as red and blue lines, respectively, and energies are referenced to the Fermi level. \textcolor{black}{The divergence of the electronic density of states at the Fermi level ($\rho \propto 1/\sqrt{E}$) stems from the one-dimensional nature of the parabolic bands ($E \propto k^2$), cf.\ Fig.\ 1(c,d).}}
    \label{Fig2}
\end{figure}

The unipolar charge transfer between $h$BN and the ZGNR is responsible for the development of the one-dimensional metallic channels and the mirror-symmetric band structures of type-I and type-II heterojunctions with respect to the Fermi level (cf.\ \textcolor{black}{Fig.}\ 1). The underlying mechanism is outlined in \textcolor{black}{Fig.}\ 2. In free-standing ZGNRs, each of the two edges hosts a pair of localized electronic states \textcolor{black}{which}, upon incorporation in $h$BN, are subject to a shift in energy. In type-I heterojunctions, the negative charge residing at the interface carbon atoms induces \textcolor{black}{an} upward energy shift, whereas in type-II heterojunctions the positive charge induces \textcolor{black}{a} downward energy shift. In both cases, the shift promotes the formation of a metallic phase. A closely related mechanism is operative in type-III heterojunctions \cite{Tepliakov2023b} as well as in free-standing ZGNRs under transverse external \cite{Son2006b} and built-in electric fields \cite{Kan2008}. The main difference \textcolor{black}{in those cases} is that the energy shift of the two pairs of edge states occurs in opposite directions due to the ambipolar charge transfer at the interfaces, \textcolor{black}{which results from} the different chemical termination of the edge carbon atoms.

\begin{figure}[t]
    \centering
\includegraphics[width=1\columnwidth]{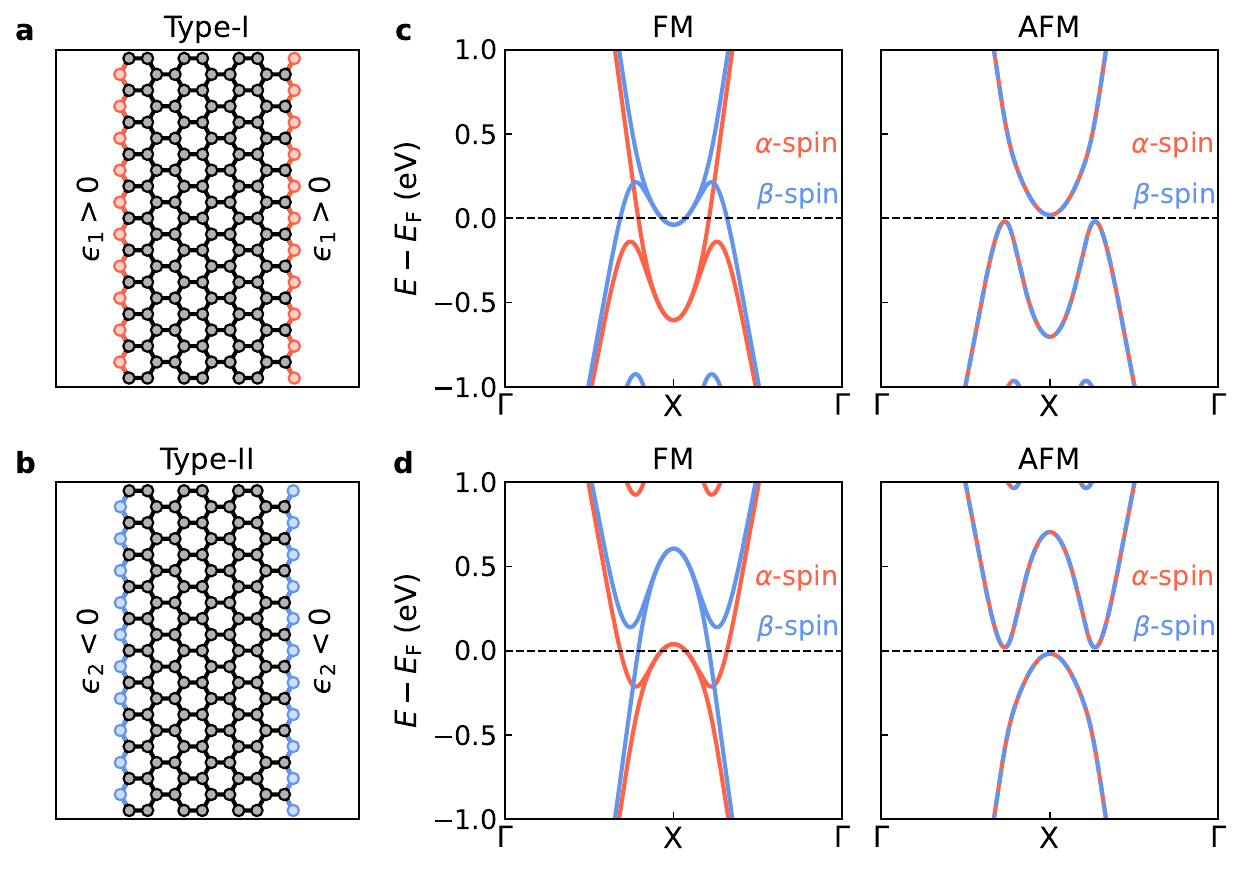}
    \caption{\textbf{Electronic structure of ZGNR/$h$BN heterojunctions from mean-field Hubbard model calculations.} 
    \textbf{a}, Lattice model of a type-I heterojunction, where the edge sites experience positive on-site energies $\epsilon_1 = |\epsilon|$.
    \textbf{b}, Lattice model of a type-II heterojunction, where the edge sites experience negative on-site energies $\epsilon_2 = -|\epsilon|$.
    In panels \textbf{a} and \textbf{c}, the edge lattice sites that experience positive and negative on-site energies are denoted as red and blue circles, respectively.
    \textbf{c}, Spin-resolved electronic band structures of a type-I heterojunction for a representative ZGNR of width 1.4 nm shown in panel \textbf{a} in the ferromagnetic (FM) and antiferromagnetic (AFM) phase.  
    \textbf{d}, Spin-resolved electronic band structures of a type-II heterojunction for a representative ZGNR of width 1.4 nm shown in panel \textbf{b} in the ferromagnetic (FM) and antiferromagnetic (AFM) phase.  
    In panels \textbf{b} and \textbf{d}, $\alpha$-spin and  $\beta$-spin electronic bands are indicated as red and blue lines, respectively, and energies are \textcolor{black}{referenced} to the Fermi level. The absolute value of the on-site potential at the edges is set to $\vert \epsilon \vert  = 0.374t$ ($t = $ 2.75 eV), as determined by fitting the \emph{ab initio} band structures.}
    \label{Fig3}
\end{figure} 

\begin{figure}[]
    \centering
\includegraphics[width=1\columnwidth]{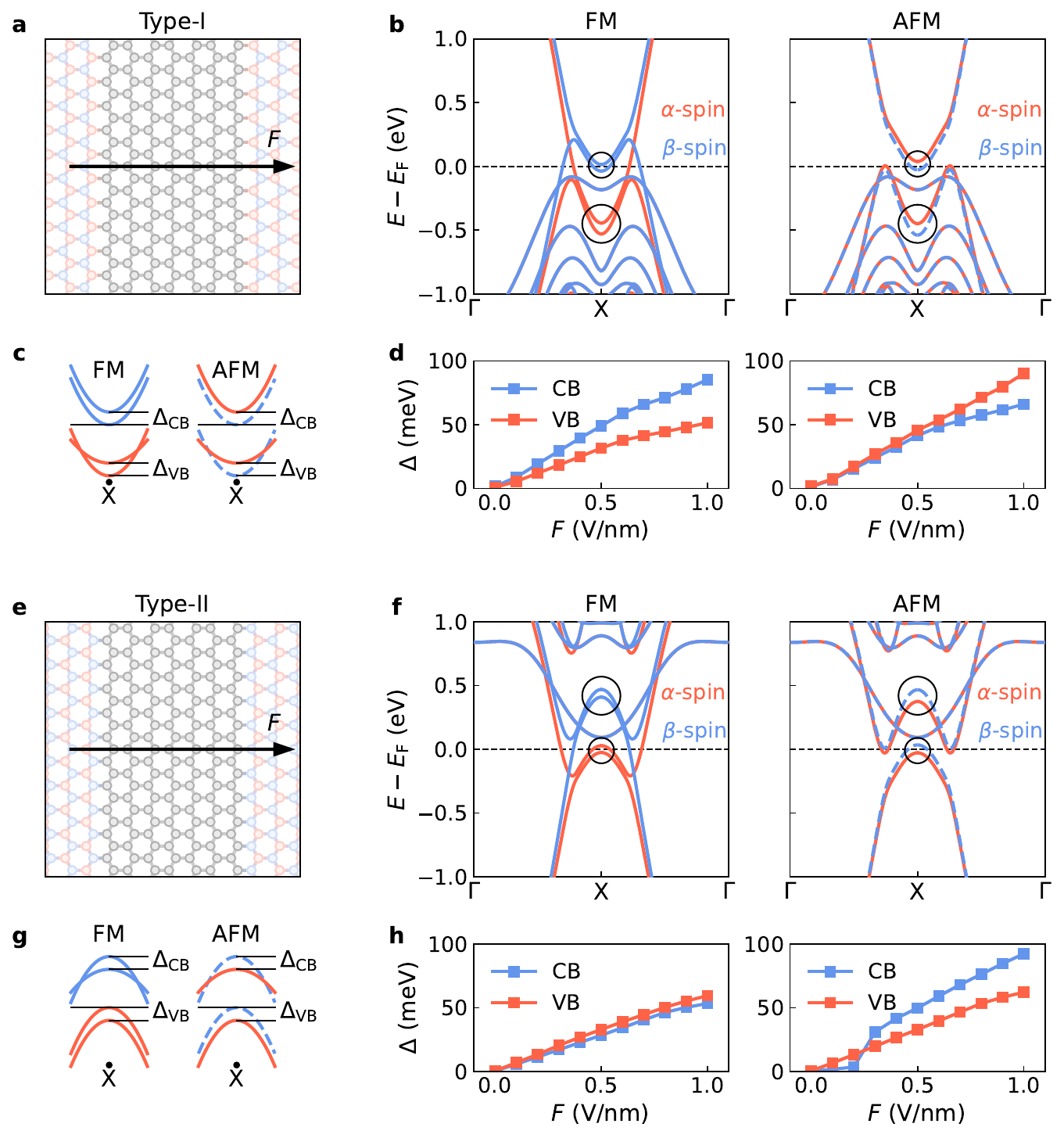}
    \caption{\textbf{Response of ZGNR/$h$BN heterojunctions to an external electric field from \emph{ab initio} calculations.} 
     \textbf{a}, Illustration of the application of a transverse external electric field $F$ to type-I heterojunctions. 
     \textbf{b}, Spin-resolved electronic band structures of a type-I heterojunction for a representative ZGNR of width  1.4 nm in the ferromagnetic (FM) and antiferromagnetic (AFM) phase at $F$ = 1 V/nm.
\textbf{c}, Illustration of the energy and spin splittings in the vicinity of the conduction band (CB) and valence band (VB) at the $X$ point of the Brillouin zone induced by the electric field (cf.\ black circles in panel \textbf{b}) in the ferromagnetic (FM) and antiferromagnetic (AFM) phases for type-I heterojunctions.
\textbf{d}, Evolution of the energy splitting $\Delta$ as a function of the field $F$ for the conduction band (CB) and valence band (VB) in the ferromagnetic (FM) and antiferromagnetic (AFM) phase for a type-I heterojunction.
     \textbf{e}, Illustration of the application of a transverse external electric field $F$ to type-II heterojunctions. 
     \textbf{f}, Spin-resolved electronic band structures of a type-II heterojunction for a representative ZGNR of width 1.4 nm  in the ferromagnetic (FM) and antiferromagnetic (AFM) phase at $F$ = 1 V/nm.
\textbf{g}, Illustration of the energy and spin splittings in the vicinity of the conduction band (CB) and valence band (VB) at the $X$ point of the Brillouin zone induced by the electric field (cf.\ black circles in panel \textbf{f}) in the ferromagnetic (FM) and antiferromagnetic (AFM) phases for type-II heterojunctions.
\textbf{h}, Evolution of the energy splitting $\Delta$, \textcolor{black}{in units of $a/\pi$ where $a$ is the lattice constant of the ZGNR,} as a function of the field $F$ for the conduction band (CB) and valence band (VB) in the ferromagnetic (FM) and antiferromagnetic (AFM) phase for a for type-II heterojunction.
In panels \textbf{b} and \textbf{f}, $\alpha$-spin and  $\beta$-spin electronic bands are indicated as red and blue lines, respectively, energies are referenced to the Fermi level.
    }
    \label{Fig4}
\end{figure}
On the basis of this \emph{ab initio} analysis of the interfacial charge transfer, we propose a simple model to capture the essential features of the electronic structure of these heterojunctions. Similar to previous works \cite{Soriano2012, Pizzochero2021a, Tepliakov2023b}, our model relies on the mean-field Hubbard \textcolor{black}{h}amiltonian and is restricted to the $p_z$-electrons of the ZGNR, without explicitly accounting for the $p_z$-electrons of $h$BN,
\begin{equation}
\begin{split}
\hat{\cal H} = -t \sum_{\langle i,j \rangle, \sigma} \left[\hat c_{i\sigma}^{\dagger} \hat c_{j\sigma} + \mathrm{h.c.}\right] + \sum_{i,\sigma} \epsilon_i \hat c_{i\sigma}^{\dagger} \hat c_{i\sigma} + \\
+ U \sum_i \left[\hat n_{i\uparrow}\langle\hat n_{i\downarrow}\rangle + \langle\hat n_{i\uparrow}\rangle\hat n_{i\downarrow}\right],
\end{split}
\end{equation}
where $\hat c_{i\sigma}$ ($\hat c_{i\sigma}^{\dagger}$) is the annihilation  (creation) operator for a $p_z$-electron with spin $\sigma$ at lattice site $i$ ($\mathrm{h.c.}$ denotes the \textcolor{black}{h}ermitian conjugate), $t$ is the hopping amplitude between nearest-neighboring lattice sites $i$ and $j$, $\epsilon_i$ is the on-site potential acting on lattice site $i$, $\hat n_{i\sigma} = \hat c_{i\sigma}^{\dagger} \hat c_{i\sigma}$ is the spin density at lattice site $i$, and $U$ is the strength of the on-site Coulomb repulsion between a pair of $p_z$-electrons located at the same lattice site. Following earlier \emph{ab initio} calculations \cite{Pisani2007} and experimentally-inferred values obtained in $sp^2$-hybridized carbon chains \cite{Thomann1985}, we set $U = t$ = 2.75 eV and use $\epsilon_i$ to control the interfacial charge transfer, as we describe below.   Further details of the solution of this mean-field Hubbard \textcolor{black}{model} are provided in Supporting Note S10.

Using \textcolor{black}{the hamiltonian defined in Eq.\ (1)}, we model the charge transfer occurring at the interfaces between ZGNR and $h$BN  in the actual heterojunctions by setting the on-site potential at the edge lattice sites to $\epsilon_1 > 0$ in type-I heterojunctions and $\epsilon_2 < 0$ in type-II heterojunctions, while $\epsilon_i = 0$ at all other sites. Our lattice models are depicted in \textcolor{black}{Fig.}\ 3(a,b). In \textcolor{black}{Fig.}\ 3(c,d), we \textcolor{black}{present} the resulting electronic band structures for the ferromagnetic and antiferromagnetic phases. Despite its simplicity, this mean-field Hubbard \textcolor{black}{model} provides band structures in excellent agreement with the \emph{ab initio} results presented in \textcolor{black}{Fig.}\ 1(c,d). Hence, the model allows us to conclude that the main effect of hexagonal boron nitride on the embedded graphene nanoribbons is the charge transfer at the interfaces and accompanying energy shift of the localized edge states. More importantly, it offers an intuitive yet accurate account of the electronic structure of these heterojunctions. \textcolor{black}{The model also describes the effect of the width of the embedded nanoribbon on the band structure of type-I and type-II heterojunctions}, as seen in Supporting Note S11,  and the Dirac half-semimetallicity of type-III heterojunctions, as discussed in an earlier work \cite{Tepliakov2023b}.

Next, we investigate the response of type-I and type-II heterojunctions to external fields, a critical issue for device integration and operation. \textcolor{black}{Through \emph{ab initio} calculations}, we consider in-plane electric fields transverse to the periodic direction of the nanoribbon, as schematically illustrated in \textcolor{black}{Fig.}\ 4(a,e),  and of strength within the experimentally accessible range. We do not observe any changes in either the relative stability of the magnetic phases or magnetic moments at the interfaces upon the application of the electric field.
The resulting band structures in the ferromagnetic and antiferromagnetic phases under a representative transverse electric field of 1 V/nm are shown in \textcolor{black}{Fig.}\ 4(b,f). The evolution of the band structures with increasing strength of the electric field is presented in Supporting Note S12. We identify two main effects. First, the electric field shifts the bands originating from $h$BN closer to the Fermi level. Second, the field-induced symmetry breaking lifts the degeneracy of the bands at the center of the Brillouin zone, as shown in \textcolor{black}{Fig.}\ 4(c,g). Depending on whether the ferromagnetic or antiferromagnetic phase is considered, an energy splitting occurs between pairs of bands possessing the same or opposite spin orientation, respectively. This effect is quantified in \textcolor{black}{Fig.}\ 4(d,h), where we show that the magnitude of these splittings increases approximately linearly with the strength of the electric field. Such a field effect is of particular interest in the metastable antiferromagnetic phase, where a spin-degenerate to spin-split transition occurs. This, in turn, leads to a reversible, electrically controlled spin-polarization of the charge carriers in the vicinity of the Fermi level.

\begin{figure}[]
    \centering
\includegraphics[width=1\columnwidth]{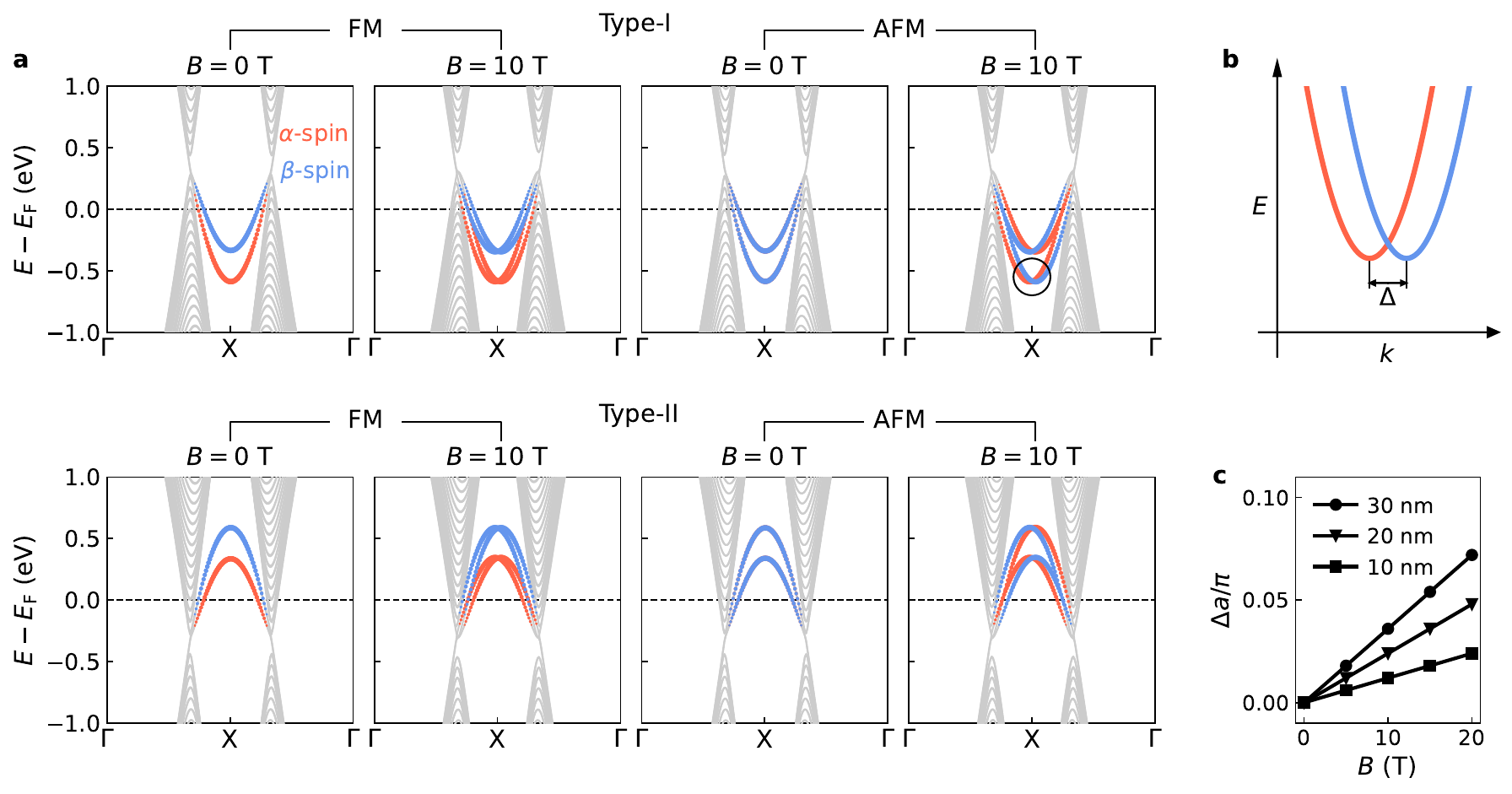}
    \caption{\textbf{Response of ZGNR/$h$BN heterojunctions to an external magnetic field from mean-field Hubbard model calculations.}  \textbf{a}, Spin-resolved electronic band structure of type-I (upper panels) and type-II (bottom panels) heterojunctions with and without an external magnetic field $B =$ 10 T in the ferromagnetic (FM) and antiferromagnetic (AFM) phase for a representative ZGNR of width 20 nm. $\alpha$-spin and  $\beta$-spin electronic bands are indicated as red and blue lines, respectively, and energies are referenced to the Fermi level. \textcolor{black}{In the AFM phase at $B =$ 0 T, the $\alpha$-spin and  $\beta$-spin electronic bands are degenerate}.  \textbf{b}, Schematic illustration of the momentum-splitting $\Delta$ induced by the external magnetic field in the heterojunctions (cf.\ black circle in panel \textbf{a}). \textbf{c}, Evolution of $\Delta$ with the external magnetic field $B$ in the heterojunctions for  \textcolor{black}{different} width\textcolor{black}{s} $w$ of the ZGNR embedded in $h$BN.
    }
    \label{Fig5}
\end{figure}

 \textcolor{black}{We can} examine the effect of an external out-of-plane magnetic field on ZGNR/$h$BN heterojunctions \textcolor{black}{using the mean-field Hubbard model of Eq.\ (1)}. We orient the vector potential of the magnetic field parallel to the periodic direction of the nanoribbon by fixing the gauge as $\vec{A} = (0, Bx, 0)$, where $B$ is the strength of the magnetic field. Since the vector potential is not dependent on the periodic coordinate $y$,  the Bloch theorem is preserved and thus we are not restricted to values of $B$ that ensure an integer number of magnetic flux quanta per unit cell. As is customary, we describe the magnetic field through  the Peierls substitution. We represent the \textcolor{black}{modified} hopping between pairs of lattice sites $i$ and $j$ as $t_{ij} = te^{i\varphi_{ij}}$, where
\begin{equation}
	\varphi_{ij} = \frac{|e|}{\hbar} \int \mathbf A \cdot d\mathbf r,
\end{equation}
\textcolor{black}{and} the integration is performed along the path connecting lattice sites $i$ and $j$. In the chosen gauge, this \textcolor{black}{gives}
\begin{equation}
    \varphi_{ij} = \frac{2\pi B}{\phi_0} \frac{x_i + x_j}{2} (y_j - y_i),
\end{equation}
with $\phi_0$ being the magnetic flux quantum while $x_i$ and $y_i$ are the Cartesian coordinates of lattice site $i$. The magnetic field introduces complex phases into the hopping integrals, breaking the time-reversal symmetry of the mean-field Hubbard \textcolor{black}{h}amiltonian, \textcolor{black}{which}  remains \textcolor{black}{h}ermitian with real eigenvalues, since $\varphi_{ij} = - \varphi_{ji}$. In \textcolor{black}{Fig.}\ 5(a), we show the band structures of type-I and type-II heterojunctions with and without an external magnetic field. We consider a relatively wide graphene nanoribbon (20 nm) to emphasize the effect of the magnetic field on the band structure. Analogous effects are observed irrespective of the width of the embedded nanoribbon, as we show in Supporting Note S13. Unlike the electric field, which lifts the energy degeneracy of the edge states, the magnetic field lifts the  momentum degeneracy, as schematically shown in \textcolor{black}{Fig.}\ 5(b). In \textcolor{black}{Fig.}\ 5(c), we quantify the resulting field-induced splitting, which is found to increase linearly with both the intensity of the magnetic field and the width of the ZGNR embedded in $h$BN.

Motivated by their recent experimental fabrication, we have investigated lateral heterojunctions of zigzag graphene nanoribbons embedded in hexagonal boron nitride. Through a combination of \emph{ab initio} and model \textcolor{black}{h}amiltonian calculations, we have focused on the hitherto unexplored symmetric configurations of the interfaces, in which both edges of the nanoribbon are terminated by the same chemical species, either boron or nitrogen atoms. We have predicted the formation of width-independent magnetic conduction channels across insulating hexagonal boron nitride. The origin of the metallic behavior traces back to the unipolar charge transfer at the interfaces between the nanoribbon and hexagonal boron nitride and the ensuing energy shift of the states residing at the edge carbon atoms. We have additionally examined the response of these heterojunctions to external electric and magnetic fields, \textcolor{black}{which can provide a} certain degree of tunability of the energy, spin, and momentum splittings in the band structure. \textcolor{black}{O}ur work establishes \textcolor{black}{that} zigzag graphene nanoribbon/hexagonal boron nitride heterojunctions \textcolor{black}{are} an \textcolor{black}{interesting} platform to explore and manipulate spin transport in the ultimate limit of miniaturization, with potential implications for spintronics and related quantum technologies.

\section{{Acknowledgments}}  
M.P.\ and N.V.T.\ contributed equally to this work. M.P.\ acknowledges support by the STC Center for Integrated Quantum Materials and the National Science Foundation (Grant No.\   DMR-1231319 and Award No.\  DMR-1922172)  N.V.T.\ acknowledges support by the President's PhD Scholarship of Imperial College London. E.K.\ acknowledges support from  the STC Center for Integrated Quantum Materials  (Grant No.\  DMR-1231319) and the Simons Foundation (Award No.\ 896626). We used computational resources from the FASRC Cannon cluster supported by the FAS Division of Science Research Computing Group at Harvard University.

\bibliography{Final-Bibliography}

\end{document}